\documentclass[10pt,letterpaper]{article}
\usepackage{geometry}

\usepackage[T1]{fontenc}
\usepackage{amsmath,amssymb}

\usepackage[utf8x]{inputenc}

\usepackage{textcomp,marvosym}

\usepackage{cite}

\usepackage{nameref,hyperref}

\usepackage[right]{lineno}

\usepackage{microtype}
\DisableLigatures[f]{encoding = *, family = * }

\usepackage[table]{xcolor}

\usepackage{array}

\newcolumntype{+}{!{\vrule width 2pt}}

\newlength\savedwidth

\usepackage[aboveskip=1pt,labelfont=bf,labelsep=period,justification=raggedright,singlelinecheck=off]{caption}

\makeatletter
\renewcommand{\@biblabel}[1]{\quad#1.}
\makeatother

\usepackage{lastpage,fancyhdr,graphicx}
\usepackage{epstopdf}
\usepackage{lipsum} % Required to insert dummy text
\usepackage[version=4]{mhchem}
\usepackage{siunitx}
\usepackage{verbatim}
\usepackage{caption}
\usepackage{subcaption}
\usepackage{amsfonts}
\usepackage{amsmath}
\usepackage{amssymb}
\usepackage{calrsfs}
\usepackage{color}
\usepackage[capitalize]{cleveref}
\usepackage{nameref}
\usepackage{graphicx}
\usepackage{media9}
\usepackage{float}
\floatstyle{plain}
\newfloat{video}{thp}{lop}
\floatname{video}{Video}
\crefname{video}{Video}{Videos}

\crefname{equation}{Eq}{Eq}

\newcommand{\benum}{\begin{enumerate}}
\newcommand{\eenum}{\end{enumerate}}

\begin{document}

\vspace*{0.2in}

{\Large
%\textbf\newline{Network analyses of 4D genome datasets automate detection of community-scale gene structure and plasticity} 
%\textbf\newline{Community-scale gene structure and plasticity in 4D genome datasets}
\textbf\newline{Transient crosslinking kinetics optimize gene cluster interactions}    
% Please use "sentence case" for title and headings (capitalize only the first word in a title (or heading), the first word in a subtitle (or subheading), and any proper nouns).
}
\newline
\\
Benjamin Walker\textsuperscript{1},
Dane Taylor\textsuperscript{2,3},
Josh Lawrimore\textsuperscript{7},
Caitlin Hult\textsuperscript{4,5},
David Adalsteinsson\textsuperscript{1},
Kerry Bloom\textsuperscript{7},
M. Gregory Forest\textsuperscript{1,6*}
\\
\bigskip
\textbf{1} Department of Mathematics, University of North Carolina at Chapel Hill
\\
\textbf{2} {Department of Mathematics, University at Buffalo, State University of New York
\\
\textbf{3} Computational and Data Enabled Science and Engineering, University at Buffalo, State University of New York
\\
\textbf{4} Department of Microbiology and Immunology, University of Michigan Medical School
\\
\textbf{5} Department of Chemical Engineering, University of Michigan
\\
\textbf{6} Departments of Applied Physical Sciences and Biomedical Engineering, University of North Carolina at Chapel Hill
\\
\textbf{7} Department of Biology, University of North Carolina at Chapel Hill
\bigskip

% Use the asterisk to denote corresponding authorship and provide email address in note below.
* forest@unc.edu

%Abstract must be 300 words or less
\begin{abstract}

Our understanding of how chromosomes structurally organize and dynamically  interact has been revolutionized through the lens of long-chain polymer physics. Major protein contributors to chromosome structure and dynamics are condensin and cohesin that stochastically generate loops within and between chains, and entrap proximal strands of sister chromatids.  In this paper, we explore the ability of transient, protein-mediated, gene-gene crosslinks to induce clusters of genes, thereby dynamic architecture, within the highly repeated ribosomal DNA that comprises the nucleolus of budding yeast. We implement three approaches: live cell microscopy; computational modeling of the full genome during G1 in budding yeast, exploring four decades of timescales for transient crosslinks between $5k bp$ domains (genes) in the nucleolus on Chromosome XII; and,  
temporal network models with automated community (cluster) detection algorithms applied to the full range of 4D modeling datasets.  The data analysis tools detect and track gene clusters, their size, number, persistence time, and their plasticity (deformation).  Of biological significance, our analysis reveals an optimal mean crosslink lifetime that promotes pairwise and cluster gene interactions through ``flexible'' clustering. In this state, large gene clusters self-assemble yet frequently interact (merge and separate), marked by gene exchanges between clusters, which in turn maximizes global gene interactions in the nucleolus.  This regime stands between two limiting cases each with far less global gene interactions: with shorter crosslink lifetimes, "rigid" clustering emerges with clusters that interact infrequently; with longer crosslink lifetimes, there is a dissolution of clusters. These observations are compared with imaging experiments on a normal yeast strain and two condensin-modified mutant cell strains. We apply the same image analysis pipeline to the experimental and simulated datasets, providing support for the modeling predictions.  

% 276 words in current abstract

\end{abstract}

\section*{Introduction}
The 4D Nucleome Project \cite{Dekker20174d}
proposes the integration of diverse approaches: increasingly powerful chromosome conformation capture techniques including high-throughput chromosome conformation capture (Hi-C); statistical and topological analyses of these massive Hi-C datasets; 3-dimensional (3D) and 4D super-resolution imaging datasets; and
%, two 
computational modeling approaches, both constrained by and independent of Hi-C datasets. The project aims to gain mechanistic understanding of 3D structure and dynamics of the genome within the nucleus, and to learn how the active chromosome architecture facilitates nuclear functions. In this paper, we contribute to these aims by combining three approaches: 
(i)~live cell microscopy for experiments studying the effect on gene clustering for normal and condensin-modified mutant cell strains;
(ii) first-principles-based, computational modeling based on the statistical physics of chromosome polymers coupled with transient gene-gene crosslinks formed by condensin proteins; 
and (iii) analysis of the dynamic chromosome architecture with temporal network community detection algorithms applied to 4D modeling datasets across four decades of crosslinking timescales.   

Our present understanding of basic principles that govern high-order genome organization can be attributed to incorporation of the physical properties of long-chain polymers \cite{marko1994fluctuations,marko1995statistical,wong2012predictive,wong2013build,vasquez2014polymer,wang2015principles}. 
The fluctuations of long-chain polymers, numerically simulated with Rouse-like bead-spring chain models of chromosomes confined to the nucleus, capture the tendency of chromosomes to self-associate and occupy territories~\cite{tjong2012physical,verdaasdonk2013centromere,ea2015contribution,vasquez2016entropy}
In addition, these models make predictions with regard to the spatial and dynamic timescales of inter-chromosomal interactions, a dynamic analog of topologically associated domains. 
The convergence of robust physical models with high-throughput biological data reveals the fractal nature of chromosome organization, namely an apparently self-similar cascade of loops within loops, or structure within structure, as one examines chromosomes at higher and higher resolution~\cite{dekker2002capturing,le2013high,rao20143d}. 

{\it De novo} stochastic bead-spring polymer models of the dynamics and conformation of ``live'' chromosomes, plus the action on top of the genome by transient binding interactions of structural maintenance of chromosome ({\it SMC}) proteins, e.g. condensin, provide complementary information to chromosome conformation capture (3C) techniques, genome-wide high-throughput (Hi-C) techniques, and restraint-based modeling \cite{dekker2002capturing,dekker2008mapping,lieberman2009comprehensive,duan2010three,tanizawa2010mapping,pope2014topologically,vian2018energetics}.  3C and Hi-C experiments rely on population averages of gene-gene proximity on all chromosomes over many thousands of dead cells whose chromosomes have been permanently crosslinked by formaldehyde; the restraint-based modeling approach then explores 3D chromosome architecture that optimizes agreement with the experimental data on gene-gene frequency and proximity across the genome. Many powerful inferences have been drawn from both Hi-C and polymer modeling approaches, using analyses of empirical and synthetic datasets encoding maps related to the pairwise distances between genes.

A common major limitation for existing polymer models and whole-genome contact maps in mammalian cells is in mapping two essential regions of the chromosome, namely the centromere and the nucleolus.  The centromere, essential for chromosome segregation, and the nucleolus, the sub-nuclear domain of ribosomal DNA, are comprised of megabases of repeated DNA (centromere satellites and nucleolus rDNA). Furthermore, these regions are not captured in methods used for generating contact maps. 
We have used single live cell imaging of the nucleolus in budding yeast coupled with whole genome polymer modeling to explore the minimal requirements for sub-compartmentalization. Implementation of protein-mediated cross-linking within the nucleolus is sufficient to partition this region of the genome from the remaining chromosomes. Furthermore, stochastic polymer models reveal that the relative timescales of crosslinking kinetics and fluctuations of the chromosome chains have a profound influence on nucleolar morphology~\cite{hult2017enrichment}.

In single cells, the positional fluctuations of tagged DNA sequences on specific chromosomes~\cite{belmont1998vivo,pearson2001budding,fisher2009dna} through the lac operator/lac repressor reporter system validate the bead-spring models. Chromosomes fluctuate as predicted for the conformational dynamics of idealized Rouse chains~\cite{belmont1998visualization}.

Polymer simulations over the entire genome have revealed the ability of relatively fast binding and unbinding, and thereby short-lived (fraction of a second) protein crosslinks to concentrate the rDNA chain sequence in a smaller volume and increase the simulated fluorescent signal intensity variance when the model datasets were convolved with a point spread function to create two-dimensional, maximum intensity projections~\cite{hult2017enrichment}. Visualizations of the monomers in the simulations revealed that the fastest kinetics explored, or shortest-lived crosslinks ($\sim.09$s), generated several clusters of high polymer density, and overall compaction of the nucleolus. In contrast, much slower kinetics (decades longer-lived protein crosslinks ($\sim90$s) ) tended to homogenize the fluorescent signal intensity as evidenced in the decrease in simulated fluorescent signal intensity variance. These model visualizations were consistent with experimental results on live budding yeast.

There is a growing interest to analyze Hi-C datasets and model chromosome interactions using network models \cite{rajapakse2010networking,rajapakse2011emerging,rajapakse2011dynamics}, which has opened the door to study chromosomal datasets using network-based algorithms including community detection \cite{xu2015hidden,chen2016spectral,cabreros2016detecting,schmitt2016compendium} and centrality \cite{liu2017genome,liu2018dynamic}.
These detection algorithms perform an unbiased search for robust structures (communities or clusters) at the scale they exist in an automated manner, quantifying how chromosome conformational changes can precede changes to transcription factors and gene expression \cite{chen2015functional,delaneau2019chromatin} and  leading to new approaches for cellular reprogramming \cite{ronquist2017algorithm,liu2017genome}. 

Here, we apply {\it temporal} community detection algorithms \cite{mucha2010community} to simulated 4D datasets over four decades of {\it SMC}-binding kinetic timescales, and thereby detect, track and label transient gene communities (clusters) in the nucleolus.  Simultaneously, we record summary statistics on the sizes and numbers as well as persistence times of communities, and the frequencies of community interactions leading to gene exchanges. We likewise record standard bead-bead summary statistics.
In doing so, we detect spatial and temporal organization at the scales they exist, beyond two-point (gene-gene) spatial proximity statistics. We identify the timescales over which spatial organization persists, linking the timescales to the cluster identification algorithm. Since clusters can deform through the flux of genes into and out of clusters, we further are able to identify crosslink timescales for which spatial clustering persists over extended timescales, and whether individual clusters are relatively permanent or experience frequent interactions and gene exchanges.  Perhaps the most striking prediction of our modeling and data analysis is that specific gene organization tasks (amplified below) are optimized at a relatively short crosslink timescale, on the order of $.19$ sec.

With these network tools applied to physics-based 4D nucleome simulated datasets, we explore the mechanistic basis for the experimentally observed variance in nucleolar morphology. {\it From a high-resolution sampling of the timescales for crosslinking of 5k base pair (bp) domains, 4D model simulations of the yeast genome reveal the nucleolus on Chromosome XII undergoes a stark transition in dynamics and structure, and does so within a narrow "mean on" crosslink timescale range of $.09 - 1.6$ sec}.
A highly stable clustering regime exists with relatively short-lived crosslinks ($~.09$ sec), with relatively few cluster interactions and gene exchanges, as reported previously in \cite{hult2017enrichment}. At slightly longer-lived ($~.19$ sec) timescales, a novel "flexible" cluster behavior is revealed. Gene clusters continue to self-organize, yet clusters are more mobile, frequently interact, and exchange genes. Indeed, {\it there is a peak timescale, marked by highly mobile gene clusters, at which both pairwise and community-scale gene interactions are maximized.}  As the binding affinity of crosslinker proteins increases only slightly longer ($~1.6$ sec), the community-scale structure has dissolved, with no identifiable nucleolar sub-substructure.  See \cref{fig:model}.

From a methods perspective, our analysis of the 4D simulated datasets is based on network modeling and a temporal community-detection algorithm known as multilayer modularity \cite{mucha2010community}. From a biological perspective, this tunable dynamic self-organization reflects a powerful mechanism to coordinate gene regulation and the coalescence of non-contiguous genes into identifiable clusters (substructures). The transition shown in \cref{fig:model} occurs {\it within} such a narrow crosslinker timescale regime ($.09 - 1.6$ sec), suggesting a relatively simple mechanism to control dynamic sub-organization of the genome; indeed we performed and report experiments below to support this prediction. Finally, we emphasize the counter-intuitive nature of this mechanism: clustering is most often associated with segregation, however we observe that the dynamic element of flexible clusters facilitates an overall increase in global gene interactions in the nucleolus.

%%%%%%%%%%%%%%%%%%%%%%%%%%%%%%%%%%%%%%%%%%%%%%%%%%%%%%%%%
%%%%%%%%%%%%%%%%%%%%%%%%%%%%%%%%%%%%%%%%%%%%%%%%%%%%%%%%%
\section*{Results}\label{sec:results}
%%%%%%%%%%%%%%%%%%%%%%%%%%%%%%%%%%%%%%%%%%%%%%%%%%%%%%%%%
%%%%%%%%%%%%%%%%%%%%%%%%%%%%%%%%%%%%%%%%%%%%%%%%%%%%%%%%%
Our results are based on analysis of the model of the yeast genome originally presented in \cite{hult2017enrichment}. An overview of the model is provided in the \nameref{sec:model} subsection in the \nameref{sec:methods} section.
We begin by extending results from \cite{hult2017enrichment} in
context with the extensions presented in this paper. In addition to simulated 4D datasets as in \cite{hult2017enrichment}, here we compare wild-type and SMC protein-altered mutant experimental data to explore how the SMC-protein crosslinking timescale $\mu$ influences behavior of the nucleolus.
In~\cite{hult2017enrichment}, three values of
$\mu \in \{0.09,{0.9},90\}$ were studied.  In this paper, we logarithmically sample across the full range of $\mu$ values between $0.09$ and $90$, providing a detailed investigation of consequences induced by the crosslinking timescale $\mu$.
We then construct simulated microscope images \cite{hult2017enrichment} for all 4D simulated datasets and compare them to experimental images of yeast nucleoli, including wild-type (WT), hmo1$\Delta$ and fob1$\Delta$.
We show that the mutations hmo1$\Delta$ and fob1$\Delta$ alter the nucleolus similarly to changes induced by varying $\mu$ in our model.

In \cite{hult2017enrichment}, 4D imaging of the nucleolus beads provided visual evidence of clusters at the shortest crosslinking timescale ($\mu \approx.09$ sec) and a lack of clusters at the longest timescales ($\mu \approx 90$ sec). Herein, we simulated across four decades of $\mu$, revealing transitions in dynamic architecture that were not explored in \cite{hult2017enrichment} since we did not have the tools to automate detection apart from obvious visual images at polar extremes, robust clusters and no clusters. Depending on the crosslinking timescale $\mu$, the nucleolar beads can aggregate into self-organized clusters, which may or may not change over time via mergers/divisions of clusters with bead exchanges between clusters.  The choice $\mu \approx 1.6$  gives rise to plasticity in the dynamic architecture of the nucleolus, whereby beads aggregate into clusters, yet the clusters frequently come into contact, merge into a super cluster, then rapidly divide with several exchanges of beads.  This process of frequent cluster interactions optimizes the traditional statistics of the frequency of pairwise bead-bead interactions within the nucleolus, while furthermore generalizing pairwise interactions to gene cluster interactions and the frequency of pairs of beads to share the same cluster. By a fine sampling of the crosslinking timescales, coupled with the automated dynamic community detection algorithms, we discover the novel non-monotone behavior in the interaction statistics, at both gene-gene and gene cluster scales.  

Motivated by the need to identify, label, and track clusters over time, we adapt a method based on network modeling and the temporal community-detection algorithm known as multilayer modularity~\cite{mucha2010community}. Communities in networks are the analog to clusters of points. Our methodology not only confirms the striking visual evidence of robust clustering at $\mu = .09$ in \cite{hult2017enrichment}, but allows us to track and label whether clusters of genes exist, cluster interactions (merger and division), their frequency, and gene exchanges per interaction. If there are no clusters, the algorithm reports that all clusters are essentially single genes with doublets and triplets that interact and deform. Our methods therefore reveal dynamic sub-organization features at the scales they exist, automatically and unbiased, as we scan 4D datasets across 4 decades of crosslinking timescales.
We show our algorithm efficiently identifies robust as well as flexible communities using a cost function that factors in both physical proximity 
and temporal coherence. Finally, we demonstrate that the identification and analysis of time-evolving communities reveals a larger scale explanation for the non-monotone behavior in bead-bead interaction statistics versus crosslinking timescales.  The explanation lies precisely in the transition from robust, non-interacting clusters for $\mu = .09$ sec, to flexible clusters for $\mu = .19$ sec, and then a slow dissolution of clusters as $\mu$ increases, with essentially no clustering by $\mu = 1.6$.

\subsection*{Transient Crosslinking Timescale Influences Nucleolus Clustering}
\label{sec:data}

We first focus in the relatively short crosslink timescale regime, extending the simulations of \cite{hult2017enrichment} at discrete values $\mu = 0.09, 0.9, 90$.
These will establish a basic understanding of how the kinetic timescale $\mu$ for crosslinking sensitively affects the organization of the nucleolus and the dynamics of the architecture.  From our refined simulations across the above four decades, the essence of the story can be told with results for three selected values $\mu\in\{0.09,0.19,1.6\}$.
In \cref{fig:model}(A--C), we present visualizations, i.e., ``snapshots,'' of the beads' 3D positions during the simulations. The nucleolus on Chromosome XII is highlighted in blue and all remaining chromosome arms are colored gray. In \cref{fig:model}(D--F), we show only the nucleolar beads, which are colored according to the network community detection analyses that we describe in the following sections. 
We also show videos of the time evolution of the beads, along with a simulated microscope projection, for each timescale in \nameref{S2_Video}, \nameref{S3_Video}, and \nameref{S4_Video}.

\begin{figure}[t!]
\centering{
\includegraphics[width=.9\textwidth]{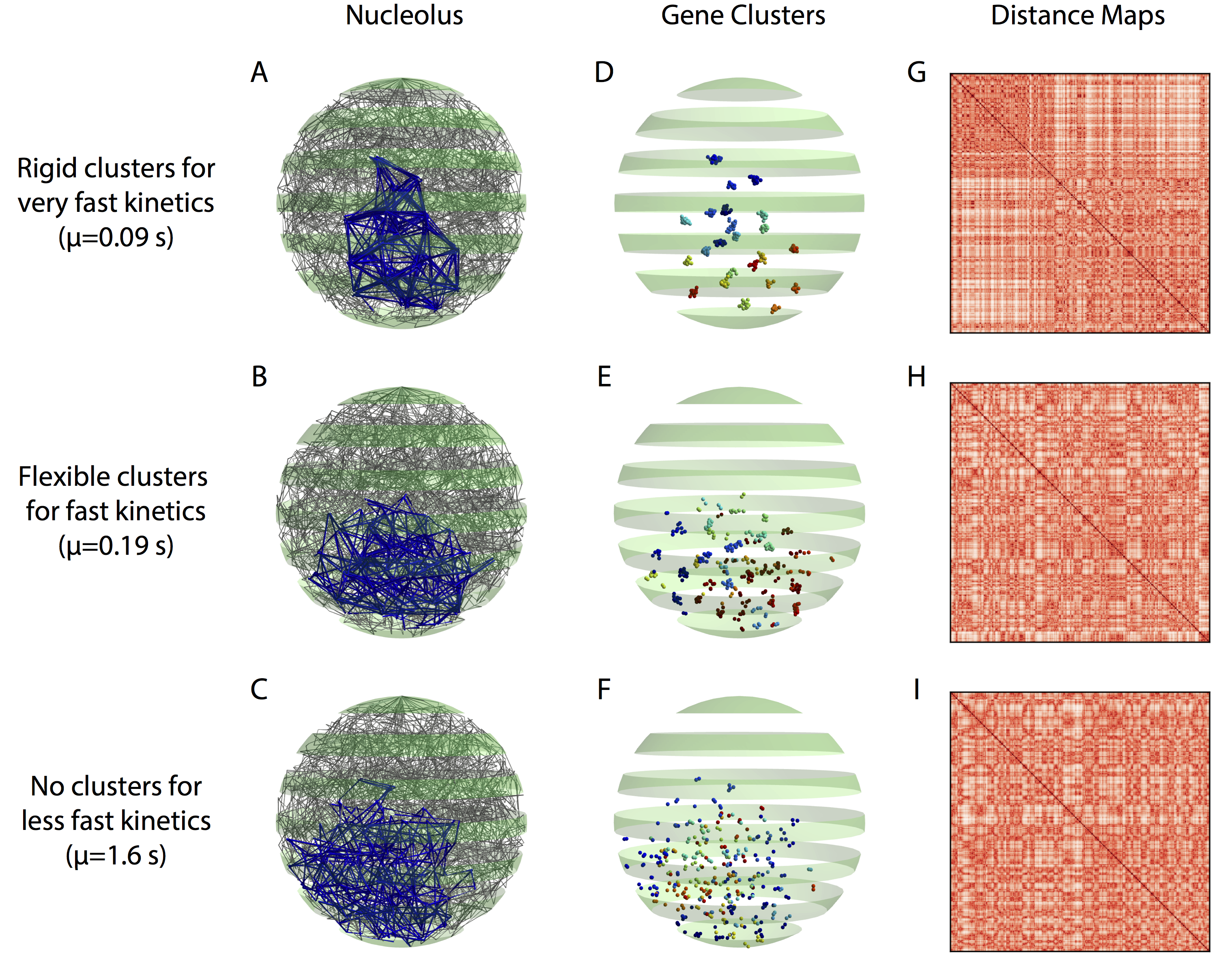}
}
\caption{Interphase yeast genome (A-C: Full, D-I: Nucleolus)
using the polymer bead-spring model of \cite{hult2017enrichment} (see Section: \nameref{sec:model}) that implements transient SMC-protein-mediated crosslinking of 5k bp domains within the nucleolus on Chromosome XII. 
The top, middle, and bottom rows depict chromosome conformations for three values of the kinetic timescale parameter $\mu$: 
(top, $\mu = 0.09$) induces rigid clusters;  
(center, ~$\mu=0.19$) induces flexible clusters; 
(bottom, $\mu=1.6$) induces no clusters. 
We identify the timescales with the intra-nucleolar clustering behavior they induce.
(A)--(C)~3D ``snapshots'' of all 16 yeast chromosomes during interphase. Blue beads and edges highlight the nucleolus.
%, which compacts as  $\mu$ decreases.
(D)--(F)~Visualization of nucleolar beads ($5k bp$ chromosome domains) that self-organize into clusters. The beads' positions are identical to those in (A)--(C) and their colors indicate their cluster labels, identified using Louvain modularity optimization \cite{newman2006modularity} as described in Methods Section: \nameref{sec:com}.
(G)--(I)~Heatmaps of the pairwise distances between beads in the nucleolus from one snapshot of the 4D time series, which provide an analogue of Hi-C bead-bead proximity data (see Methods Section: \nameref{sec:meth_distance}). Note that it is difficult to predict the absence/presence of clusters from heat maps.
}
\label{fig:model}
\end{figure}

Based on \cref{fig:model}(A)--(F) and the videos, we identify three qualitative regimes for nucleolus clustering:
\begin{enumerate}
\item \textit{rigid clustering} whereby strong, stable clusters arise, e.g., with $\mu=0.09$. 
\item  \textit{flexible clustering, or cluster plasticity} with
slightly weaker clustering whereby communities of genes form and persist, however the clusters frequently interact, merge, and divide, swapping genes per interaction, e.g., with $\mu=0.19$. 
\item \textit{non-clustering} with a lack of robust communities in which the beads act as lone units, pairs or triplets, e.g., with $\mu=1.6$.
\end{enumerate}
We will continue to use this terminology when referring to these three clustering regimes.
We note that \cite{hult2017enrichment} discovered the two extreme regimes: robust clusters for $\mu=0.09$, and the lack of clusters for $\mu=90$. As they did not finely sample the decades of timescales in between, they did not discover that the transition from robust to no clustering is in fact non-monotone with respect to gene-gene interactions, nor that the transition is essentially complete already at $\mu = 1.6$, and that the most biologically interesting and relevant regime occurs at $\mu=0.19$. Furthermore, without automated structure detection algorithms, they would not have been able to detect and dynamically track clusters of genes and their interactions that explain the peak in gene-gene interactions at $\mu=0.19$. This transition behavior and the optimal properties that arise will be the focus of several sections to follow.

In \cref{fig:model}(G)--(I), we show heatmaps of the bead-bead distances associated with the bead positions of the snapshots in (D)--(F), identical to those in (A)--(C); construction of heatmaps is described in the Methods Section: \nameref{sec:meth_distance}. Heatmaps are widely used in Hi-C to depict population averages of pairwise gene-gene proximity data ~\cite{eser2017,schalbetter2017,fudenberg2016loopextrusion,dekker2013exploring,imakaev2012,lieberman2009comprehensive} and in simulated data from polymer bead-spring models, both from 3D snapshots and time averages  \cite{verdaasdonk2013centromere,vasquez2016entropy,hult2017enrichment}.  
Comparing the second and third columns of Fig.~\ref{fig:model}, we note the difficulty (false negatives and false positives) in detecting the presence of structure and sub-organization in column 2 from visual examination of heatmaps in column 3. 

As shown in \cite{hult2017enrichment}, the time average of 4D simulated datasets, even in the strong clustering regime, wipes out the sub-structure of snapshots when averaging over the entire G1 phase. An alternative approach has been to use polymer modeling to generate chromosome conformations, and to select those conformations that best match Hi-C data, so-called restraint-based polymer modeling \cite{Dekker20174d}. 
Simultaneously, there have been efforts to develop methodologies to identify gene clusters in a rigorous and automated way from Hi-C data \cite{rajapakse2010networking,rajapakse2011emerging,rajapakse2011dynamics}.  
Our conclusion is that there is a need for a more reliable and objective method to study the clustering of chromosome domains in the nucleolus, especially spatio-temporal methods that take into account how bead positions and sub-organization change with time, weighing both spatial proximity and temporal coherence in the detection method. In the following sections, we present a scalable and automated technique to identify and track the dynamics of clusters. First, however, we will present new experiments that provide empirical evidence for clustering in the nucleolus.

\subsection*{Evidence of Nucleolus Clustering in Experimental and Simulated Microscopy Images}
\label{sec:res_exp}
We conducted experiments to compare our 
simulations to empirical measurements obtained from live cell microscopy, extending the results previously reported in \cite{hult2017enrichment}.
Here, we study three yeast strains: wild-type (WT), fob1 and hmo1. Importantly, fob1$\Delta$ and hmo1$\Delta$ are mutations that lack
key proteins reported to crosslink or loop segments of rDNA within the nucleolus. 
Fob1$\Delta$ is required for maintenance of the rDNA copy number and regulates the association of condensin with rDNA repeats~\cite{johzuka2009cis,johzuka2006condensin}. The replication fork barrier within the rDNA is a binding site for Fob1$\Delta$ that, together with several other components (Tof1, Csm1 and Lrs4), are responsible for the concentration of condensin within the nucleolus~\cite{johzuka2009cis}. Strain hmo1 is an abundant high mobility group protein that localizes to the nucleolus and has been proposed to share functions with UBF1, which is involved in rDNA transcriptional regulation within the nucleolus~\cite{prieto2007recruitment,albert2013structure}. Fob1 and Hmo1 are non-essential genes and were deleted from the genome to allow us to study their effect on nucleolus morphology due to functional modifications of crosslinking. See Methods Section: \nameref{subsec:strain} for further  details.

\begin{figure}
    \centering
    \includegraphics[width=\textwidth]{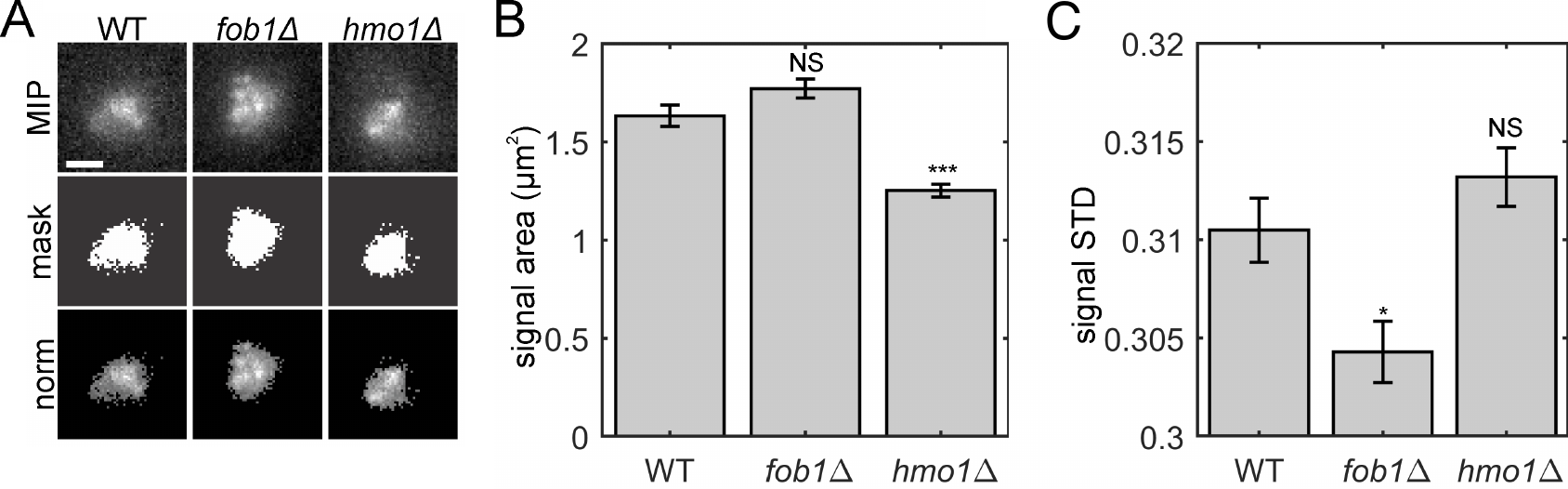}
    \caption{
    Experimental images of the nucleolus.  Algorithmic thresholding of CDC14-GFP reveals alterations in nucleolar area and variance of signal intensity.
    (A)~Top row depicts images of maximum intensity projection of WT, fob1$\Delta$, and hmo1$\Delta$ yeast cells containing CDC14-GFP to label the nucleolus. Scale bar is 1  $\mu$m.  
    The middle row are images of the binary mask generated by thresholding the projections using Otsu's method. The bottom row are images of the projections where the intensities of pixels below the threshold were set to zero. See Methods Section: \nameref{subsec:image} for more detail.
    (B)~Bar chart of the nucleolar signal are calculated by the area of the binary mask. 
    (C)~Bar chart of the standard deviation of normalized nucleolar signal. 
   In panels (B) and (C), the non-significant changes are labeled `NS' (see text), the bars represent an average value across $n$ cells, and error bars indicate standard error.
    }
    \label{fig:bio_fig1}
\end{figure}

In \cref{fig:bio_fig1}, we present images and analyses of nucleoli  of these strains using fluorescent, live-cell microscopy.
To visualize nucleoli, we fused Cdc14 protein phosphatase to green fluorescent protein (GFP) \cite{hult2017enrichment}. Nucleolar protein fusions occupy a distinct region of the nucleus that is adjacent to the nuclear envelope and (typically) opposed to the spindle pole body. 
We describe the image acquisition and processing steps in Methods Section: \nameref{subsec:image}, and we highlight a few details here.
Following image acquisition, we construct \emph{maximum intensity projections} (MIP) centered on the nucleolus. See top row of \cref{fig:bio_fig1}(A).}
Due to potential variation in CDC14-GFP protein copy number and nucleolar/rDNA size from cell to cell, we normalized the nucleolar CDC14-GFP signal after excluding all intensity values below a intensity threshold. To this end, we first selected a threshold using Otsu's  method~\cite{otsu1979threshold}, which we implemented using the MATLAB function multithresh. One can interpret the threshold as a binary mask, as shown in the second row of \cref{fig:bio_fig1}(A). After applying the mask,
we normalized the nucleolar signal by subtracting all intensities by the minimum value and then dividing them 
by the new maximum intensity that is obtained after subtraction.
The third row of \cref{fig:bio_fig1}(A) depicts normalized images.
Mutations of Hmo1 and Fob1 were found to alter the area and signal intensity of nucleoli labeled with CDC14-GFP across a range of intensity thresholds, which we surmise is due to alterations in the architecture of, i.e., clustering within, the nucleolus.

In \cref{fig:bio_fig1}(B)-(C), we provide results for an analysis of  nucleolar morphology: (B) the area of nucleolar signal; and (C) the standard deviation of the normalized signal. This analysis was implemented using 
the numerical algorithms presented in \cite{hult2017enrichment}, which we further describe in Methods Section: \nameref{subsec:image2}.

As shown in \cref{fig:bio_fig1}(B),
null mutations of hmo1$\Delta$ significantly altered the area of the nucleolar signal, whereas null mutations of fob1$\Delta$ did not. This was assessed by a Student's two-tailed T-test, which yielded 
$p = 3\times10^{-8}$ for the former and $p = 0.07$ for the latter.
As shown in \cref{fig:bio_fig1}(C), the standard deviations of the normalized images were
significantly lowered for the fob1$\Delta$ null mutation, but this did not occur for the hmo1$\Delta$ null mutation $p = 0.01$ versus $p = 0.2$).
The non-significant changes are labeled `NS' in the figure. The error bars indicate standard errors across $n$ cells, where $n=84$, $70$ and $77$ for the WT, fob1$\Delta$ and hmo1$\Delta$ strains, respectively.

We  note that \cite{hult2017enrichment} also studied the area and variance of the nucleolus using experimental and simulated images. They found, for example,
that the distribution of areas occupied by the nucleolus displays a lognormal distribution for WT cells in G1. Also, recall that we implemented thresholding based on Otsu's method; in contrast, \cite{hult2017enrichment} explored a range of threshold values and found qualitatively similar results to be consistent across a range of threshold values. They did not, however, explore the area and variance for simulated images for a wide range of $\mu$, which is the focus of our next experiment.

\begin{figure}
    \centering{
    \includegraphics[width=\textwidth]{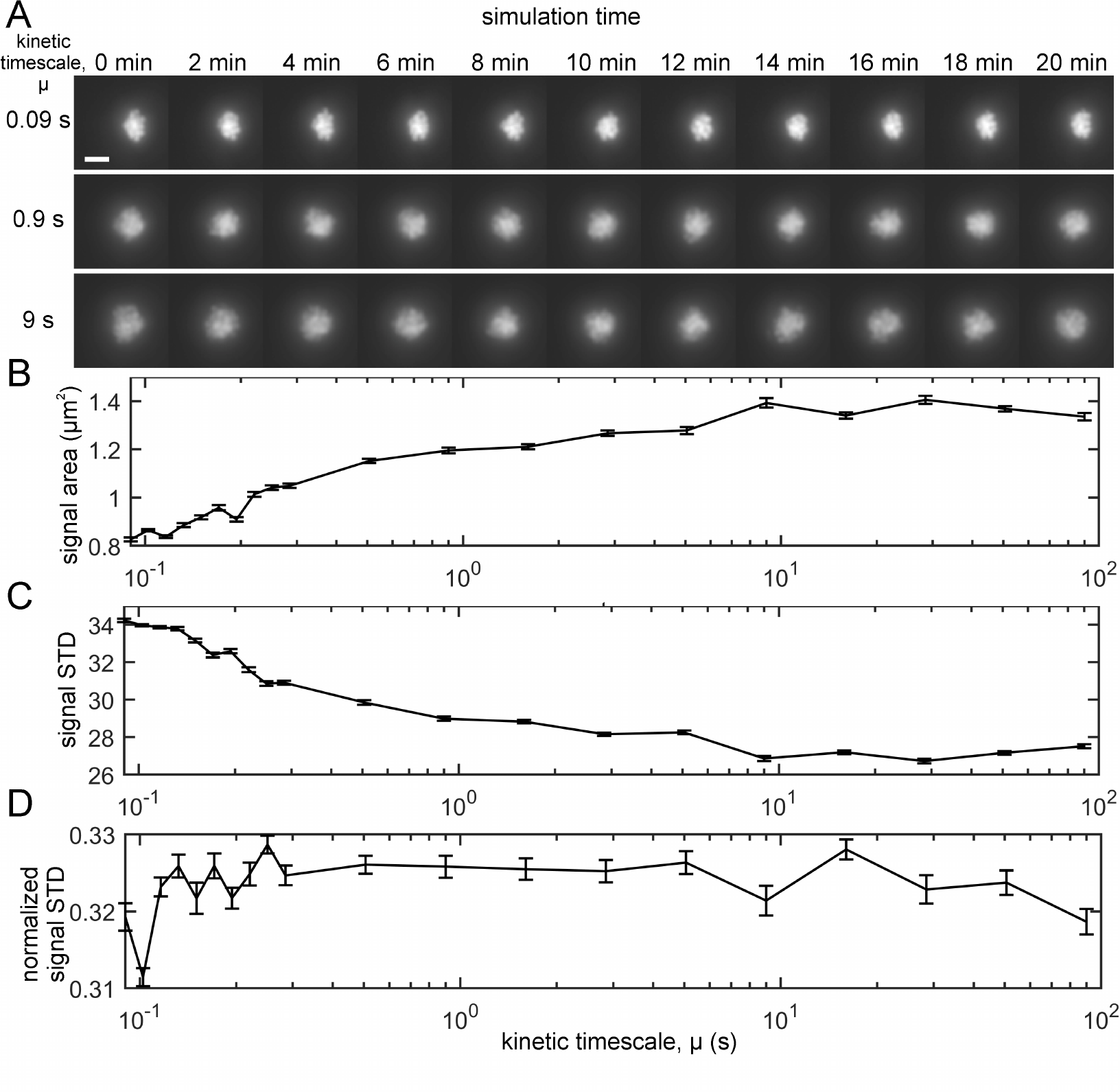}
    }
    \caption{Simulated microscope images for polymer bead-spring model with transient crosslinking and varying kinetic timescale $\mu$.
    Varying $\mu$ alters the area and variance of the intensity for the nucleolar signal, which models the affect of the fob1$\Delta$ and hmo1$\Delta$ mutations.
    (A)~Timelapse montage of simulated microscope images
    for $\mu\in\{0.09,0.9,0\}$ (seconds).
    Scale bar is $\mu$m. 
    (B)~Area of nucleolus signal ($\mu m^2$) 
    as a function of $\mu\in[10^{-1},10^{2}]$.
    Areas were calculated by measuring the area of the binary mask generated by applying Otsu's threshold to each image.
    (C)~Standard deviation of the (non-normalized) nucleolus signal versus $\mu$.
    (D) Standard deviation of the normalized nucleolus signals versus $\mu$; these we normalized identically to the normalization of the experimental microscope images.
    In panels (B)-(D), error bars indicate standard errors observed using 22 time points for each value of $\mu$.}
    \label{fig:bio_fig2}
\end{figure}

\begin{figure}
    \centering
    \includegraphics[width=\textwidth]{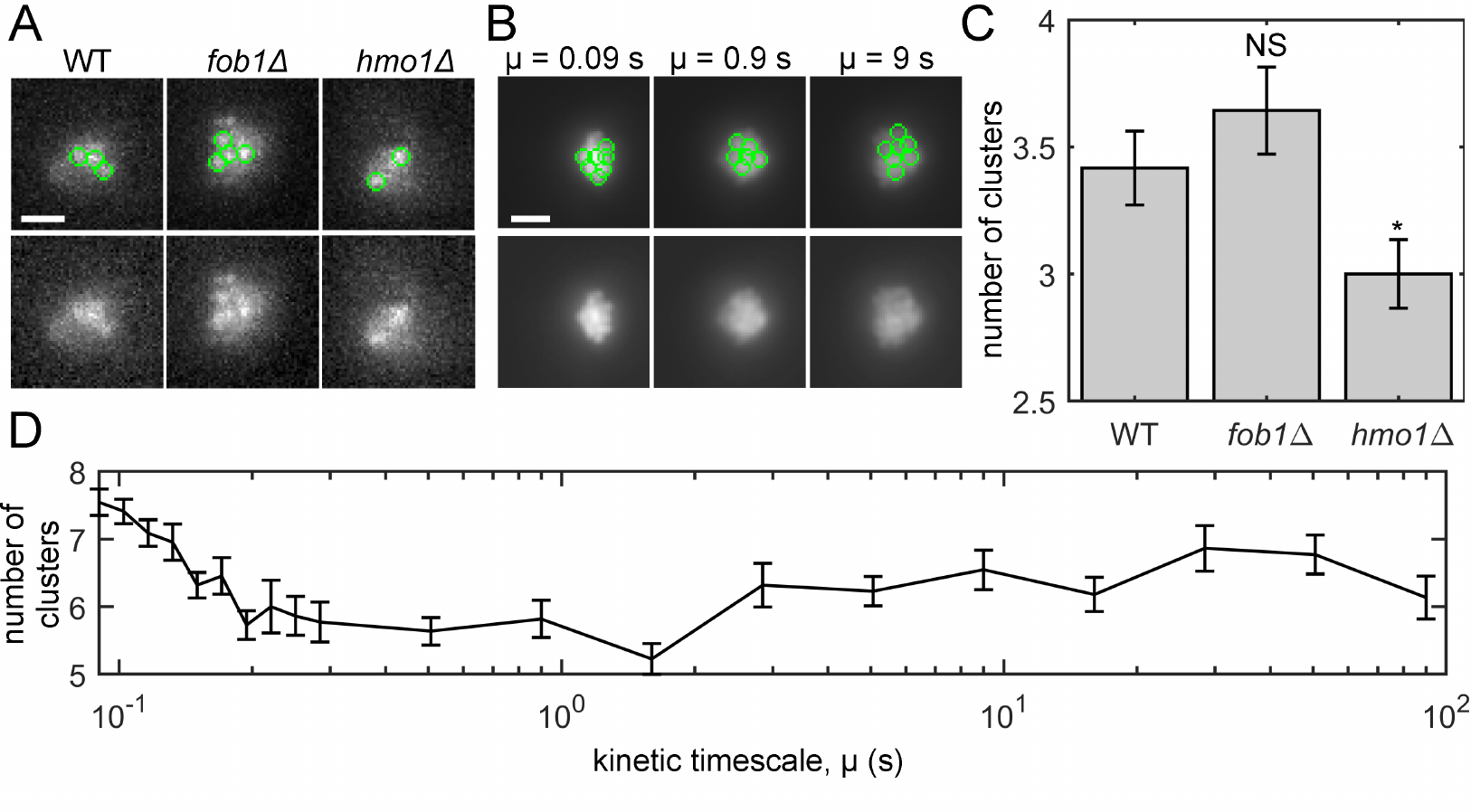}
    \caption{
    Comparison of experimental and simulated microscope images.
    Varying $\mu$ in the polymer bead-spring model captures the effect on nucleolus clustering incurred by the crosslink-altering mutations.
    (A) Maximum intensity projections of CDC14-GFP in WT, fob1$\Delta$, and hmo1$\Delta$ cells with identified clusters marked by green circles. The lower row depicts the same images as the top row, except the circles are removed.
    (Images are the same as \cref{fig:bio_fig1}(A).) 
    (B) Same information as panel (A), except the results depict simulated images based on 4D simulation data for three values of $\mu$. 
    Scale bars in (A) and (B) are 1 $\mu m$.
    (C) Bar chart showing the number of clusters for each strain. The bars and error bars indicate the average and standard error across $n$ cells, where   $n = 84$, $70$, and $77$ for WT, fob1$\Delta$, and hmo1$\Delta$, respectively. Significance was assessed via a Student's two-tailed T-test:  
    $p = 0.3$ for WT versus fob1$\Delta$ and  $p = 0.04$ for WT versus hmo1$\Delta$.
    (D) Average number of clusters for simulated images as a function of $\mu$. The averages and standard errors (see error bars) were calculated using 22 time points for each $\mu$.
    }
    \label{fig:bio_fig3}
\end{figure}

To explore whether varying the kinetic timescale $\mu$ for our simulations yields similar changes as those arising under  the fob1$\Delta$ and hmo1$\Delta$ mutations, we applied the microscope simulator of \cite{hult2017enrichment} to our 4D simulated data and analyzed the images using the same image analyses as described in \cref{fig:bio_fig1}. First, we converted our 4D simulated data into a timelapse sequence with 22 time points, i.e., snapshots.  Each nucleolus bead was convolved with a point spread function and a maximum intensity projection was created for each timepoint. We depict 11 such images in \cref{fig:bio_fig2}(A). In panel (B), we plot the area of the nucleolar signal (computed using Otsu's threshold) versus $\mu$. Note that the nucleolus area increases as $\mu$ increases.
In panel (C), we plot the standard deviation of nucleolar signal versus $\mu$, which has the opposite trend.
In \cref{fig:bio_fig2}(D), we plot the standard deviation of images obtained after a normalization step that is identical to that implemented for the experimental images (see discussion for \cref{fig:bio_fig1}.
Interestingly, the dependence on $\mu$ of the signal's standard deviation drastically changes depending on whether or not it is normalized.
Given that normalization is required to control for cell-to-cell differences in CDC14-GFP and in nucleolar/rDNA size, we sought develop a metric to measure clustering in the CDC14-GFP signal that was independent of the absolute values of the intensities.

Our final experiment studies cluster formation in the nucleolus and compares clustering observed in the experimental and simulated microscopy images.
We developed a cluster detection algorithm written with MATLAB (see Methods Section: \nameref{subsec:image2})  and applied it to both the experimental and simulated images. We have made the code available at \cite{bloom_code}.
In \cref{fig:bio_fig2}(A), we depict images of  maximum intensity projections for WT, fob1$\Delta$ and hmo1$\Delta$ strains with (top row) and without (bottom row) visualizations of detected clusters, which are represented by green circles.
In panel (B), we depict identical information as in panel (A) except we show simulated images for three values of $\mu$.
In \cref{fig:bio_fig2}(C) and (D), we show the number of clusters for the experimental and simulated images, respectively. For the experimental images, we give results for WT, fob1$\Delta$ and hmo1$\Delta$, whereas for the simulated images we present results for $\mu\in[.09,90]$.
We observe that the number of  clusters was significantly decreased in the hmo1$\Delta$ null mutation, but not the fob1$\Delta$ null mutation ($p = 0.04$ for hmo1$\Delta$ versus $p = 0.3$ for fob1$\Delta$).
We also observe that increasing $\mu$ yielded a general trend in which there were fewer clusters.
Taken together, these data suggest that gene clustering can directly impact the size and shape of the nucleolus.

\subsection*{Histograms of 2-Point Pairwise Distances between Nucleolar Beads}\label{sec:mixing}

A simple and previously used method for analyzing distances between beads is to create a histogram of all bead-bead pairwise distances. As explored in \cite{hult2017enrichment}, this two-point statistic can provide evidence of clustering and can be used to query simple properties such as whether or not the clusters change over time. In this section, we repeat this analysis on our data and extend it by showing what effects averaging over time and averaging over populations has on the results. We show that averaging one cell over time prevents observing the flexible clustering through pairwise distances, and averaging over populations prevents observing any sort of clustering. We provide further details on the computation of these distances in Methods Section: \nameref{sec:meth_distance}.

\begin{figure}[h!]
\centering
\includegraphics[width=\textwidth]{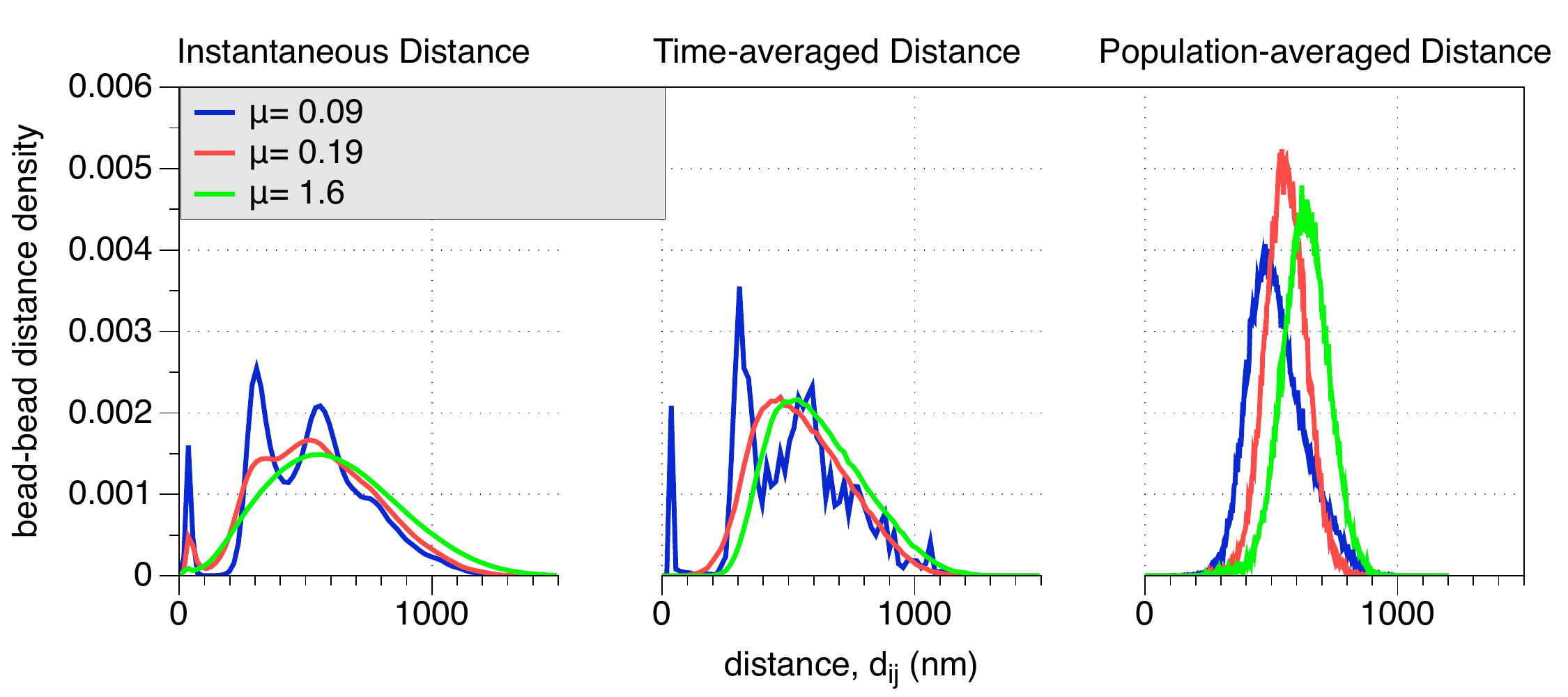}
\caption{
(A) Instantaneous distances, (B) time-average distances, and (C) population-averaged distances. In each panel, we show results for
the three kinetic regimes illustrated in \cref{fig:model}: $\mu=0.09$, $\mu=0.19$), and $\mu=1.6$. 
As reported in \cite{hult2017enrichment}, multimodal histograms are a ``signal'' for the presence of clusters. This signal is strongest for  $\mu=0.09$, is nonexistent for $\mu=1.6$, and we observe a new regime for $\mu=0.19$ (flexible clustering or cluster plasticity), whereby `soft' clusters form but deform over time through cluster interactions. The vertical dashed lines in (D)-(F) indicate a good distance threshold to construct gene-interaction networks; beads that are closer than $d^*=100nm$ are  likely to belong to the same cluster.
 }
\label{fig:pairwise_1}
\end{figure}

% Talk about the multiscale structure seen in the fast crosslinking
In \cref{fig:pairwise_1}(A), we plot the distribution of all pairwise distances $\{d_{ij}^{(t)}\}$ at a single time $t$ for three kinetic timescales, given by the same values $\mu=0.09,0.19,1.6$ as shown in \cref{fig:model}. For $\mu=0.09$, the pairwise distance distribution is clearly a multimodal distribution~\cite{hult2017enrichment}. The peak near $d \approx 50$ represents a large number of very short pairwise distances between beads in the same cluster. For slightly larger $d$, the density drops to zeros, indicating a separation distance between clusters. 
Interestingly, we  observe two more peaks near  $d \approx 300$ and $d \approx 600$. The clarity of these peaks suggests that the clusters themselves are regularly spaced from one another, reminiscent of a lattice structure. This shows the three layers of the multiscale structure of the nucleolus for $\mu=0.09$: its existence as a dense, secluded section of the nucleus, the self-organization of intra-nucleolar clusters, and the individual beads within each cluster.
For $\mu=0.19$, one can also observe in \cref{fig:pairwise_1}(A) three peaks in the empirical probability density for bead-bead distances, but these peaks are much less pronounced. This shows a gradual transition in the degree of clustering as we increase $\mu$. There is also a smaller gap between peaks. Together, these observations recapitulate our observations in \cref{fig:model}(E), wherein the clusters can be observed to be less compact.
Finally, for $\mu=1.6$, there is no multimodal structure in the bead-bead distance plot. This is consistent with our expectation that there is no clustering structure present for this range of $\mu$.

 % Figure showing the variation of structure with mean on and also with high averaging time

% Demonstrate that different mean on times lead to different cluster persistance
The rigid and flexible clustering cases differ not only in how strong the clustering is at any given time, but also in how stable the structure is in time. 
We investigate this by considering how averaging pairwise-distances either across across time (\cref{fig:pairwise_1}(B)) or over multiple simulations (\cref{fig:pairwise_1}(C)) influences pairwise-distance probability densities.

In \cref{fig:pairwise_1}(B), we plot the empirical probability densities for pairwise distances averaged across our 20 minute simulations. Note for $\mu=0.19$ that the density is no longer multimodal, implying that aggregating the data across a large time range inhibits the detection of flexible clusters, which by definition change with time. Note that the rigid clusters, which are very stable across time, remain discernable as the pairwise probability density remains multimodal. 
Unsurprisingly, the slow crosslinking appears qualitatively very similar in the long time average, as there was no apparent structure in the first place.

In \cref{fig:pairwise_1}(C), we plot the empirical probability densities for pairwise distances at a single time but averaged across 10 simulations with different random initial conditions. Note for all $\mu$ that there is no longer any multimodal structure for these densities, highlighting that averaging across heterogeneous cell populations obscures the detection of clusters.

%%%%%%%%%%%%%%%%%%%%%%%%%%%%%%%%%%%%%%%%%%%%%%%%%%%%%%%%%
\subsection*{Flexible Clustering Regime Maximizes Bead Mixing}
\label{sec:control}
%%%%%%%%%%%%%%%%%%%%%%%%%%%%%%%%%%%%%%%%%%%%%%%%%%%%%%%%%

%\mgf{I said this a bit differently in the Introduction.  Kerry did not like the term "mixing", which I favored.  So I used "interactions" which can be interpreted as communication, which Dane favored as I recall.   I also pointed out and emphasized that "maximizing pairwise gene interactions is the tip of the iceberg", since the peak likewise corresponds to a larger scale effect, of minimizing the waiting time for beads to return to the same cluster, not just to get next to one another. This happens precisely because the fast but not too fast timescales make the clusters more mobile, therefore clusters interact more frequently, swapping genes as the merged clusters then divide.       So I wonder if Dane can edit the material below with attention to this more general, larger scale interaction  }

%\todo{As written, the scope is intentionally focused on the small scale. This section observes properties at a pairwise scale, and we hypothesize that they arise as an artifact of the clustering. That is, we observe clusters at a large scale, and we observe optimality at a small scale, but until we study/analyze the clusters we cannot cannot conclude that it is the clusters that is the mechanism responsible for the observed optimality. The experiments in this motivate us to develop clustering algorithms (next section). Then we study mixing at the community level, and indeed our hypothesis is correct (next next section). }

Next, we study how the kinetic time scale $\mu$ (i.e., and thus the presence of clusters) affects the properties of pairwise \emph{gene interactions}. A pair of beads is said to be interacting if they are in very close proximity and the   distance between them drops below $d^*$. As discussed in \cref{fig:pairwise_1}, we choose $d^* = 100 nm$ unless otherwise noted.
%To this end, we introduce and study the properties of  \emph{gene-interaction networks}. As described in Methods Section: \nameref{sec:meth_network}, we define a gene interaction as an event in which two beads come in close proximity, particularly their distance is less than $d^*$, and we encode these interactions as a time-varying  network.
%
%Our network analyses rely on the distances between bead pairs, and so we first study these distances and as well as the effects of averaging the distances across a time window or across multiple cells (e.g., multiple simulations). 
%
In the following experiment, we show that increasing $\mu$ not only inhibits the formation of clusters,
but that there exists a particular range of $\mu$ that optimizes
\emph{gene mixing}, or the overall interaction frequency of all pairs of genes.
These experiments illustrate how clustering --- which inherently describes multi-way relationships --- can be studied through pairwise distances --- which inherently describe two-way relationships ---, and how there remain important open problems related to the time series signal processing of 4D chromosome conformation datasets.   
%We describe our computation of such distances in the Methods Section \nameref{sec:meth_distance}.

We study the following summary statistics for gene mixing:

\begin{itemize}
\item[(A)] The \emph{interaction fraction} indicates the fraction of possible unique bead pairs that interact at least once during an interphase simulation. 
\item[(B)] The \emph{mean interaction number} indicates the number of simultaneous interactions (i.e., beads within distance $d^*$) for a nucleolus bead, averaged across time and across beads.
\item[(C)] The \emph{mean waiting time} indicates for any two beads, selected at random, the average time that passes between their $i$-th and $(i+1)$-th interactions.
\item[(D)] The \emph{mean interaction duration} indicates the amount of time beads enter and reside within the interaction distance.
\end{itemize}

\begin{figure}[t!]
\centering
\includegraphics[width=\textwidth]{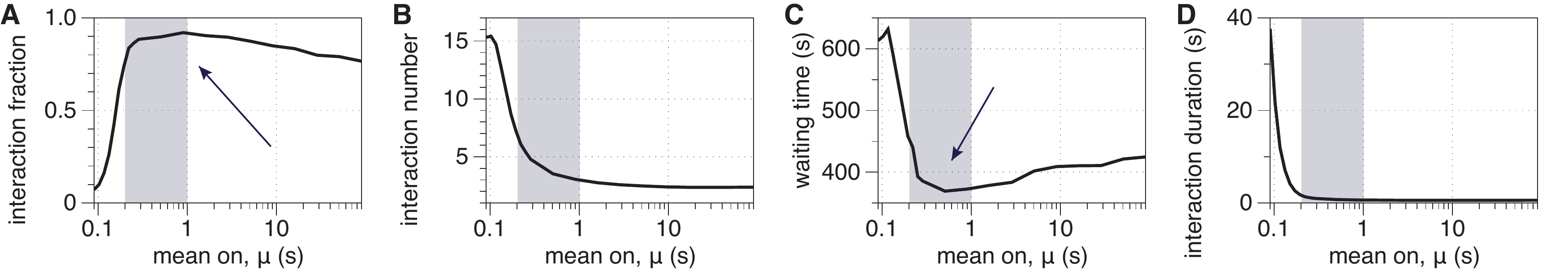}
\caption{
Mixing statistics for gene interactions:
(A) mixing fraction; (B) mean interaction number;  (C) mean waiting time; and (D) mean interaction duration, which we plot as versus  $\mu$. The shaded regions indicate the regime of flexible clustering, $\mu\in(0.19,1)$. The arrows in  panels (A) and (C) highlight that the interaction fraction and mean waiting time are both optimized for this range of $\mu$. In contrast, this regime is associated with sharp transitions  for the mean interaction number and duration, which both monotonically  decrease with $\mu$.
}
\label{fig:mixing_stats}
\end{figure}

In \cref{fig:mixing_stats}(A--D), we plot these summary statistics across a wide range of $\mu$. We identify three regimes that optimize different attributes.  $\mu\approx 0.1$ yields a self-organized structure that maximizes the number and the duration of gene-gene interactions (see panels (B) and (D)).
Recall from Video~1 that $\mu = 0.09$ yields many large clusters that are stable (i.e., do not change) over time. This is reflected in a high number of interactions with beads in the same cluster and low number of  interactions with beads not in the same cluster.

With $0.15 \lessapprox \mu \lessapprox 1$, we see flexible clustering behavior from Video~2. Notably, we find here that this flexible clustering has interesting properties beyond simply being a weaker version of the strong clustering from the rigid clustering regime. Namely, \cref{fig:mixing_stats}(A) shows that these $\mu$ values maximize the fraction of pairs of beads that interact at least once over the simulation, and \cref{fig:mixing_stats}(C) shows that these values minimize the waiting times between subsequent interactions.
Thus, we can say that \textit {flexible clustering promotes the number of both simultaneous and overall distinct pairwise gene interactions in the nucleolus}. This behavior arises from a balance between the number of intra-cluster gene-gen interactions, which is still elevated due to the moderate clustering as shown in (B), and the ability for genes to frequently switch between clusters during cluster interactions, as indicated by the reduced waiting time in (C).
SMC proteins with such crosslinking timescales  will thereby promote collective interactions among all active genes. These circumstances could accelerate a homology search, for example, to facilitate DNA repair, if the sister chromosomes were suddenly activated by a family of SMC proteins whose binding affinity was near this "sweet spot".

Finally, $\mu \gtrapprox 1.5$ is associated with a non-clustering regime, as shown in Video~3. The lack of clustering is reflected by a low number of gene-gene interactions, and the freely diffusing nature of the beads is reflected by short interaction duration and high interaction fraction.

\subsection*{Identifying Cluster Membership with Network Community Detection Algorithms}

Having found that flexible clustering  maximizes interesting properties of gene interaction, we seek to develop tools to  identify and label the spatiotemporal clusters. In the rigid clustering regime, the clusters are so well-defined that any reasonable algorithm will detect them, but this is not the case for the flexible clustering.
To detect and track flexible clustering, we utilize both spatial and temporal information to identify and track clusters.

While we have access to 4D bead position time series data, we begin by transforming this into a multilayer network problem as described in Methods Section: \nameref{sec:meth_network}. This is motivated by the fact that the most similar data available in biology, the Hi-C dataset, does not measure true distances between genome regions, but rather a notion of similarity based on average proximity~\cite{dekker2002capturing}. The result of this transformation is a time sequence of weighted, undirected networks whose edge weights represent how near two beads tend to be to each other at that point in time. We refer to this sequence of networks as a temporal temporal network.

Given a gene interaction network, we identify communities using the multilayer modularity~\cite{mucha2010community}, which we numerically solve using the Louvain algorithm \cite{blondel2008fast}. 
The algorithm, which we describe in Methods Section: \nameref{sec:multilayer_modularity},
automatically computes the communities at all time instances simultaneously, and the rate at which communities remain constant over time or are allowed to change can be tuned with parameters.
Specifically, the multilayer modularity involves the selection of two parameters $\gamma$ and $\omega$, which  are ``tuning knobs'' to control the clusters' size and temporal coherence, respectively. 
%In the Supplementary Material: S1 Text,  we describe a study in which we explore a wide range of the $(\gamma,\omega)$ parameter space and use a variety of techniques to identify appropriate parameter choices.

In \nameref{S5_Video}, \nameref{S6_Video}, and \nameref{S7_Video}, we present equivalent videos to \nameref{S2_Video}, \nameref{S3_Video}, and \nameref{S4_Video}, respectively, except in the new videos we color the beads according to the  community labels that we detect. 
These new videos provided qualitative evidence supporting the ability of our algorithm to find clusters that have appropriate spatial and temporal scales. Looking at the rigid clustering in \nameref{S5_Video}, we see the coloring strongly agrees with our visual perception in the clusters. A similar but less decisive conclusion can be made from observing \nameref{S6_Video}. 

These videos indicate good agreement between visual perception of clusters and the clusters that are detected by the multilayer modularity algorithm - when beads visually appear to be clumped together, they tend to also be the same color in the videos, which reaffirms the validity of our choice of clustering algorithm.
However, especially when looking at \nameref{S7_Video} depicting the non-clustering regime with slow crosslinking, we identify a key and common issue with clustering algorithms - they typically identify the ``best'' clusters, even when no clusters actually exist. 
%To account for this, in the Supplementary Material: S1 Text, we describe how one can distinguish between clusters that are significant and those that are not.

\subsection*{Gene Mixing at the Community Level Further Supports Flexible Clustering as the Mechanism for Optimality}
In this section, we provide further evidence that flexible clustering is the mechanism that is responsible for the 
optimality observed in  \cref{fig:mixing_stats}. To this end, we will revisit and modify our definitions for \emph{gene interactions} and \emph{gene mixing}, which were defined at the ``bead level'' (i.e., for pairs of beads). We now define similar, but slightly different, concepts that are defined at the  ``community-level'' in that they reflect only community-membership information and do not require the precise bead locations.
%
%We defined a gene interaction as an event in which two  beads are within distance $d^*$ of one another. This gave rise to our definition of a gene-interaction network and network properties that we called gene mixing statistics. 
We say that two beads are ``communicating'' if they are in the same community.
That is, all beads in the same cluster are communicating with each other, and beads in different clusters are not communicating.
With this modified definition in hand, we define summary statistics for gene mixing at the community level that are analogous to the 2-point summary statistics for pairwise gene interactions that we previously defined in Section:~\nameref{sec:control}. {Analogous to gene mixing at the bead level, we now define ``cross communication'' at the community level.}

\begin{enumerate}
    \item[(A)] The \emph{communicating fraction} indicates the fraction of bead pairs that are in the same community at least once during a simulation.
    \item[(B)] The \emph{average beads per community} indicates, for a nucleolus bead, the average number of beads in the same community at the same time, averaged across time and across beads.
    \item[(C)] The \emph{mean waiting time} indicates for any two beads, selected at random, the average time that passes between when they are no longer in the same community and when they are next in the same community.
    \item[(D)] The \emph{mean interaction duration} indicates the amount of time between beads when they are first in the same community and when they are no longer in the same community.
\end{enumerate}

\begin{figure}[h!]
\centering{
\includegraphics[width=\textwidth]{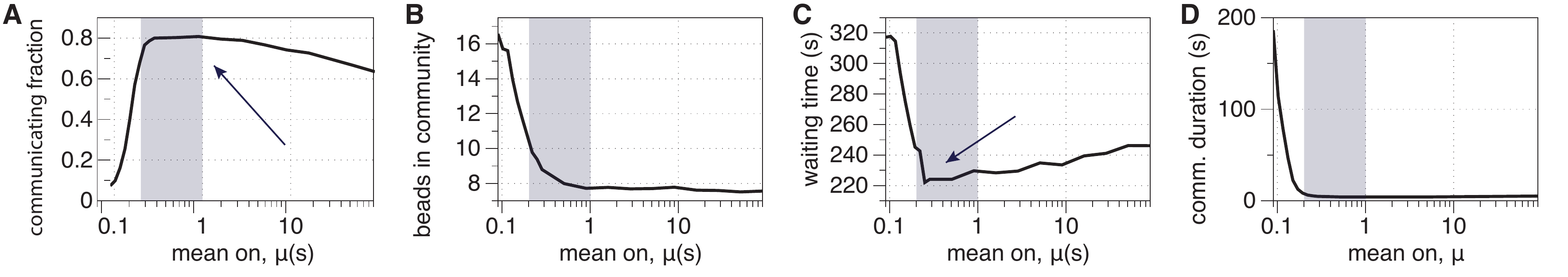}
}
\caption{
Cross communication describes the dynamics of community memberships of beads through community-level mixing.
We study cross-communication for a large range of $\mu$ by plotting four summary statistics: (A) communicating fraction; (B) mean interaction number; (C) mean waiting time; and (D) mean interaction duration (see text). The shaded regions indicate the regime of flexible clustering, $\mu\in(0.19,1)$, and the arrows in panels (A) and (C) highlight that the interaction fraction and mean waiting time are both optimized for this range of $\mu$. 
These results are qualitatively identical to the results in \cref{fig:mixing_stats} for gene mixing, illustrating that cluster formation and the exchanges of beads between clusters determines the timescale of mixing.
}
\label{fig:crosscommun}
%\end{fullwidth}
\end{figure}

In \cref{fig:crosscommun}, we present summary statistics for cross communication that are analogous to our results in \cref{fig:mixing_stats} for pairwise gene interactions. Note that the results in \cref{fig:crosscommun} are qualitatively identical to those in \cref{fig:mixing_stats}, supporting our hypothesis that gene-level mixing is determined by community-level cross communication. That is, the formation of gene clusters and exchanges of genes between them governs the timescale at which nucleolar domains come in close proximity of one another.

\subsection*{Temporal Stability of Clusters}\label{sec:stability}
Using temporal community detection, we are now finally able to quantitatively support our first observations made in Section: \nameref{sec:data} that there are three distinct clustering regimes: rigid clustering; flexible clustering; and no clustering. We support these observations by studying the properties of the detected clusters.
%We detect the clusters using temporal network models and temporal community detection.

\begin{figure}[t]
\centering{
\includegraphics[width=\textwidth]{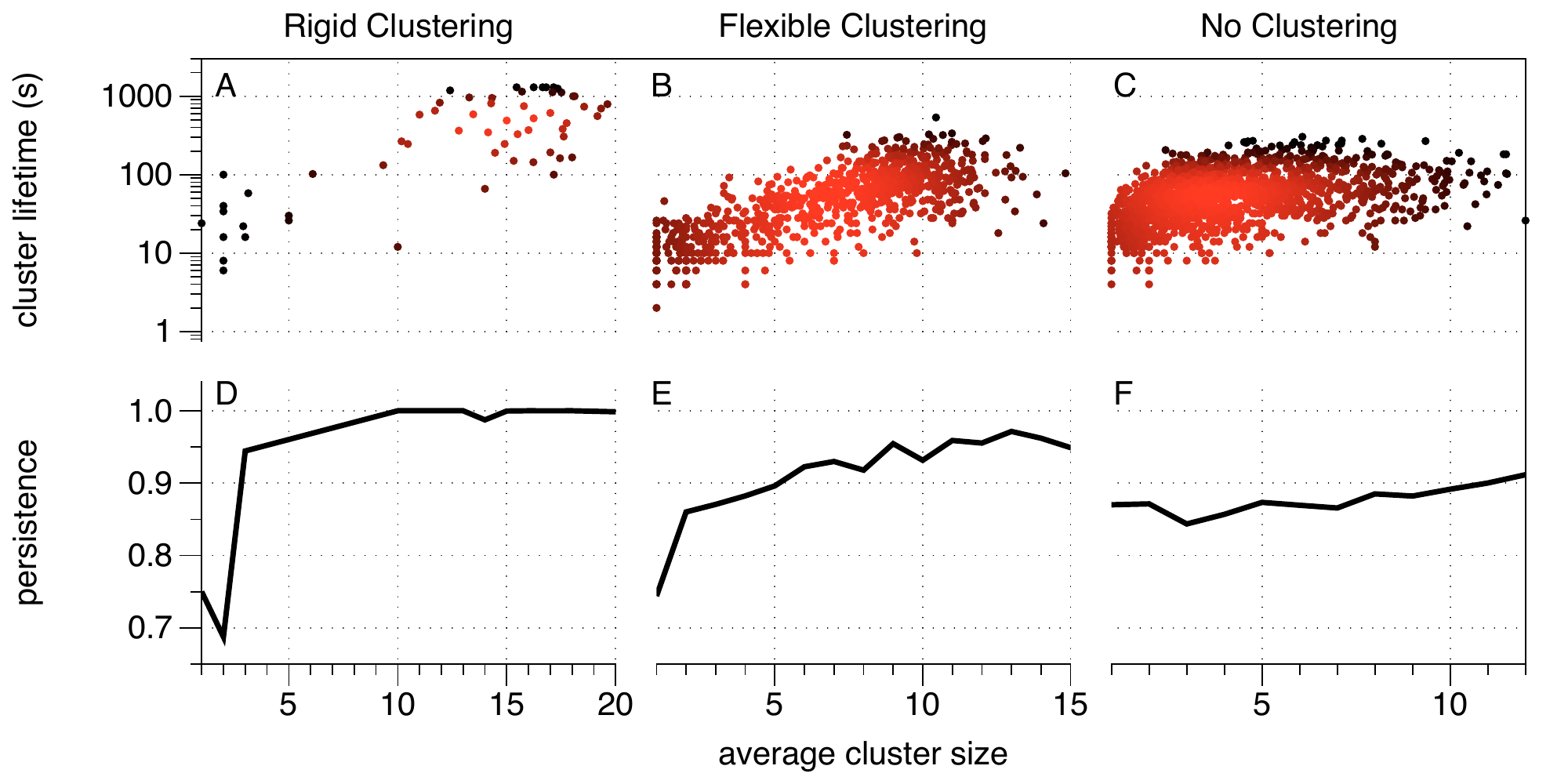}
}
\caption{
The birth and death of gene clusters identified using modularity-based community detection in temporal networks \cite{mucha2010community}.
Panels (A), (B) and (C) depict the lifetime (i.e., duration) of each gene cluster versus the average cluster size for $\mu=0.09$, $0.19$, and $1.6$, respectively. The points' colors have been chosen to highlight the density of points (with red indicating where there are many points close to one another). 
(D)--(F): The persistence (i.e., temporal coherence) of clusters is indicated by the probability that a randomly selected bead remains in the same clusters in the next time window, again plotted versus the average number of beads in that clusters. 
Results reflect $d^* = 325$, $\gamma = 10$, $\omega = 1$.
}
\label{fig:persistence1}
\end{figure}

In \cref{fig:persistence1}(A)--(C), we plot the average lifetime of clusters as a function of their average size (averaged over time). Panels (A)--(C) indicate the three clustering regimes with $\mu\in\{0.09,0.19,1.6\}$, respectively. 

For the rigid clustering regime, in \cref{fig:persistence1}(A), we see that most of the clusters are large, with an average size of 10 or more beads, and also have a long lifetime of over 100 seconds. This is consistent with our prior observations (e.g. Video 1) that showed large clusters that appeared very stable in time. We also see that the large clusters survive for much longer than the small clusters.

For the flexible clustering regime, in \cref{fig:persistence1}(B), we can see the same general trend that larger clusters tend to have a longer lifetime than smaller clusters, but there is a much wider spread of cluster sizes, with only a moderate number of large clusters.

For the non-clustering regime, there appears to be little relationship between cluster size and stability beyond an average size of approximately 3 beads. The clusters also tend to be much smaller, with almost no clusters with an average size over 10 beads.

In \cref{fig:persistence1}(D)--(E), we plot the probability that a bead remains in the same community upon the next timestep, again as a function of cluster size. Panels (D)--(E) indicate results for $\mu\in\{0.09,0.19,1.6\}$, respectively. In agreement with panels (A)--(C), one can observe that larger clusters are more stable. Note also that the communities exhibit more plasticity for $\mu=0.19 $ than for $\mu=0.09$ since beads have a higher average probability for changing the community to which they belong.

%%%%%%%%%%%%%%%%%%%%%%%%%%%%%%%%%%%%%%%%%%%%%%%%%%%%%%
%%%%%%%%%%%%%%%%%%%%%%%%%%%%%%%%%%%%%%%%%%%%%%%%%%%%%%
\section*{Discussion}
%%%%%%%%%%%%%%%%%%%%%%%%%%%%%%%%%%%%%%%%%%%%%%%%%%%%%%
%%%%%%%%%%%%%%%%%%%%%%%%%%%%%%%%%%%%%%%%%%%%%%%%%%%%%%

The dynamic self-organization of the eukaryote genome is fundamental to the understanding of life at the cellular level.  The last quarter century has witnessed remarkable technological advances that provide massive datasets of both the spatial conformation of chromosomal DNA from cell populations (3C and Hi-C generalizations) and the dynamic motion of domains in living cells (GFP tagging and tracking of specific DNA sequences), from the yeast to the human genome.  Data mining of this massive data has likewise witnessed remarkable advances in understanding the hierarchical packaging mechanisms of DNA that act on top of the genome, e.g., histones and structural maintenance of chromosome (SMC) proteins, the topology of individual chromosome fibers, their topologically associated domains, and the territories they occupy in the nucleus. The third wave of advances has come from 4D modeling of chromosomes based on stochastic models of entropic, confined polymers, and the coupling of SMC proteins that either bind and crosslink genes on chromosomes or generate loops on individual chromosomes.  As these three approaches continue to mature and inform one another, at an ever-increasing pace, insights into the structure and dynamics of the genome continue to deepen. 

The motivation for this paper lies in the information that can be inferred from these massive datasets, from Hi-C, live cell imaging experiments, and polymer physics modeling. In previous studies, cf.\cite{hult2017enrichment} and references therein, we showed that heterogeneity in experimental images derive from substructures that are formed within the nucleolus. We decided to use the array of tools from network community detection analysis, modeling, and associated fast algorithms, to automate the search for dynamic architectures in the 4D datasets generated by polymer modeling.  We note a similar network analysis approach has been applied to Hi-C datasets \cite{rajapakse2010networking,rajapakse2011emerging,rajapakse2011dynamics}, whereas our datasets have the added feature of highly resolved temporal information.  Our aim was to infer organization beyond 2-point, time-averaged or population-averaged, gene-gene proximity statistics and heat maps generated from the statistics, and to remove the bias of an individual's visual determination of structure. To do so, we used the advances in network-based models, their temporal generalization, data analysis, and algorithms, and applied this arsenal of tools to 4D datasets across four decades of crosslinking timescales to:  
(i) robustly identify clusters, or communities, of genes (5k bp domains, or beads, in our model), i.e., to directly detect gene sub-organization at the scale it exists rather than attempt to seek larger scale organization from gene-gene statistics and heat maps; 
(ii) determine the size distribution (number of genes) in such sub-structures; 
(iii) determine the persistence times of communities; and, 
(iv) determine the interaction frequency of communities and corresponding gene exchanges, which are the drivers of gene-gene interaction statistics. 
In this way, network algorithms automate gene community detection and persistence, with robustness built in by enforcing insensitivity to algorithm tuning parameters.

We elected to build and implement these network tools on the 4D datasets generated in house, from simulations of our recent polymer modeling of interphase budding yeast \cite{hult2017enrichment}.  In this model, a pool of SMC proteins transiently and indiscriminantly crosslink 5k bp domains within the nucleolus on Chromosome XII. The kinetics of the cross-linking anchors relative to the substrate is a major driver of sub-nuclear organization. If the crosslinkers bind and release more rapidly than the chains can relocate, the non-intuitive consequence is that the chains explore less space. When dense clusters of crosslinker/binding sites arise, they persist for extended time periods when the crosslinking kinetics is sufficiently fast.

We previously showed in  \cite{hult2017enrichment} that very short-lived ($\mu=0.09 sec$) binding kinetics provided closer agreement with experimental results (highest degree of compaction of the nucleolus into a crescent shape against the nuclear wall).  It was also shown via visualization of the simulated 4D datasets that this timescale induces a decomposition of the nucleolus into a large number of clusters each consisting of many 5k bp domains, and these clusters were persistent over time.   On the other hand, with long-lived crosslinks ($\mu=90 sec$) the clusters disappeared.  These results reveal that the timescales of the crosslinkers relative to entropic fluctuations of the chromosome polymer chains are a fundamental contributor to genome organization.

For the present paper, the sample set of binding kinetics in \cite{hult2017enrichment} was expanded to 4D datasets of interphase, sampling over four decades of crosslinking duration timescales.  We applied standard, distance-based, 2-point statistical metrics and visualization tools, and then analyzed the full range of 4D datasets with the fast, automated network models and tools.  The network algorithms search for and detect time-varying communities (clusters, sub-structures) at the scales they exist, not at a prescribed scale of two or more genes; the spatial and temporal scales are identified with the criteria that they are robust to the algorithm parameters.
We then use this information to label and color-code communities, using community-level description to understand the persistence and interactions (merger and division marked by gene exchanges) between communities. 

With the above community-scale information and statistics, we generalize standard gene-gene interaction statistics across the four decades of bond duration timescale.  As a generalization of waiting times for 2 distant genes beads to come within a specific distance of one another, we calculate waiting times for genes in the same community to leave and then re-enter another common community, and calculate the fraction of all genes that were in the same community at least once during interphase, which we call the community cross-communication fraction.  

From these analyses, we discovered a novel dynamic self-organization regime, wherein the rigid, persistent communities at relatively short-lived crosslink timescales ($\mu=.09 sec$) transition at slightly longer-lived crosslink timescales ($\mu=0.19 sec$) to more mobile (literally, the clusters diffuse faster) communities that interact far more frequently, each interaction corresponding to merger, subsequent division, and an exchange of genes.  We refer to this regime as flexible community structure with enhanced cross-communication.   Furthermore, we discovered non-monotonicity in the dynamic self-organization behavior: the community cross-communication fraction is maximized, coincident with a minimum waiting time between genes departing and returning to common communities, with a  crosslink timescale of $\mu=0.19 sec$.  Both properties fall off, albeit in different ways, for shorter and longer timescales.

We emphasize that these network tools and fast algorithms are amenable to any 4D dataset from polymer models.  While we restricted the analysis in this study to the nucleolus during G1 of budding yeast where SMC proteins are allowed to transiently crosslink 5k bp domains, the same analyses can be applied to data with tandem SMC crosslinking and loop generation, for any cell type and for any phase of the cell cycle.  Moreover, given the growing interest in network-based analyses for Hi-C data, network modeling is well-positioned to provide a fruitful direction for data assimilation efforts aimed at connecting simulated and empirical 4D chromosome conformation data. An important challenge facing this pursuit is the development of improved data pre-processing and community-detection methodology for temporal and multimodal network datasets \cite{taylor2016enhanced,taylor2017super}.

%%%%%%%%%%%%%%%%%%%%%%%%%%%%%%%%%%%%%%%%%%%%%%%%
%%%%%%%%%%%%%%%%%%%%%%%%%%%%%%%%%%%%%%%%%%%%%%%%
\section*{Materials and Methods}\label{sec:methods}

\subsection*{Model}
\label{sec:model}
Chromatin dynamics within confined yeast nuclei 
has been widely modeled using Rouse polymer bead-spring chains; see earlier citations. We employ the identical model and code in \cite{hult2017enrichment}, resolving the entire yeast genome into 2803 total beads, each of which represents approximately 5k base pairs (bp). As such, each bead is interpreted as a chromosome domain or a tension blob, as explained in the polymer physics literature. The beads are arranged on 32 chromosome arms having lengths that reflect their experimentally identified lengths.  Each  arm is tethered at both ends to the nuclear wall: all emanating from the centromere at one end, with the other end tethered to one of six telomeres. Along each arm, there are entropic, nonlinear
springs (the wormlike chain model is used here) that connect neighboring beads. 
The beads also experience Brownian, entropic, repulsive and hydrodynamic drag forces, and are physically confined to the nucleus. 
We simulate approximately 20 minutes of G1 during interphase, and each simulation is initialized with 32 chromosome arms tethered at the centromere and one of six telomeres on the nuclear wall, otherwise randomly located within the nucleus (idealized as a spherical domain).

We model the effect of SMC proteins by transiently crosslinking pairs of non-neighboring beads in the nucleolus, represented by a contiguous chain of 361 beads on chromosome XII. (We note in passing that \cite{hult2017enrichment} also studied, experimentally and computationally, when the nucleolus is split onto separate chromosome arms, showing this to have a negligible affect on the clustering and interaction results.)
A transient crosslink is modeled by a pair of stochastic events --- a binding and unbinding of two non-adjacent 
nucleolar beads --- and the timescale of these events 
is tuned by a single  parameter $\mu$ (measured in seconds).
We defer to \cite{hult2017enrichment} for the details, but elaborate here on the role of $\mu$. 
To implement crosslinks, all nucleolar beads are assigned a state of ``active'' or ``inactive,'' and crosslinks are  allowed only between active beads. Each bead's state fluctuates stochastically as follows: an active bead becomes inactive after a duration that is a random number drawn the normal distribution $N(\mu, (\mu/5)^2)$; and an inactive bead becomes active after a duration  drawn from $N(\mu/9, (\mu/45)^2)$. 
If two nucleolar beads are both active and the distance between them is less than $\tilde{d}=90$ nm, then a crosslink is formed between them (i.e., they bind). Each bead may be involved in at most one crosslink at a time, decided on the basis of pairwise proximity among all active beads at each timestep. If either bead becomes inactive, then the crosslink is broken (i.e., they unbind). Hence, crosslinks are established and broken stochastically in the nucleolus, and the single parameter $\mu$ dictates the kinetic timescales for crosslinking.
See \cite{hult2017enrichment} for additional model details.

\subsection*{Pairwise-Distance Maps for High-Throughput Chromosome Conformation Capture (Hi-C)}
\label{sec:meth_distance}

Each simulation of the Rouse-like polymer model yields time-series data $\{  {\bf x}^{i} (t)\} \in \mathbb{R}^3$ that defines the 3D location of each bead $i\in\{1,\dots,N\}$ at each discrete time step $t=0,1,\dots$. We establish a connection between our simulated data and the state-of-the-art in  chromosome imaging---namely, high-throughput conformation capture (Hi-C) --- by constructing and analyzing \emph{pairwise-distance maps}. Hi-C ``images'' the conformation of chromosomes using a combination of proximity-based ligation and massively parallel sequencing,  which yields a map that is correlated with the pairwise distances between gene segments. While the actual pairwise distances between gene segments cannot be directly measured, Hi-C implements spatially constrained ligation followed by a locus-specific polymerase chain reaction to obtain \emph{pairwise count maps} that are correlated with spatial proximity: the count between two gene segments monotonically decreases as the physical 3-dimensional distance between them increases.

To provide an analogue to Hi-C imaging, we construct pairwise-distance maps for our simulated data $\{  {\bf x}^{i} (t)\} $. Let 
\begin{equation}
F: \{{\bf x}^{i}\}_{i=1}^{i=N} \mapsto \mathbb{R}^{N\times N}\label{eq:dist_map}
\end{equation}
define a \emph{map} (used here in the mathematical sense) from a set of $N$ points $\{{\bf x}^{i}\}\in\mathbb{R}^3$ to a matrix such that each entry $(i,j)$ in the matrix gives the distance between point $i$ and point $j$. Whereas Hi-C imaging aims to study the positioning of chromosomes using noisy measurements that are inversely correlated with pairwise distances, for our simulations we have access to the complete information about the chromosome positioning. We therefore define and study several variations for pairwise distance maps, which will allow us to also study  artifacts that can arise under different preprocessing techniques, such as averaging the time series data across time windows and/or averaging across multiple simulations with different initial conditions.
We define the following pairwise distance maps:
\begin {itemize}
\item An \emph{instantaneous} pairwise distance map $X(t)= F(\{{\bf x}^i(t)\})$ encodes pairwise distances between  beads (i.e., chromosome domains) at a particular timestep $t$.
\item A \emph{time-averaged} pairwise distance map $Y(\tau) = \frac{1}{|\tau|} \sum_{t\in\tau} X(t)$ encodes the  pairwise distance between beads averaged across a set of timesteps $\tau$. 
\item A \emph{population-averaged} pairwise distance map $Z(t) = \langle X(t)\rangle_p$ encodes the pairwise distance at timestep $t$ between beads, which are averaged across several simulations that have different initial conditions (which are chosen uniformly at random).
\end{itemize}
These pairwise distance maps represent the data that is sought after, but cannot be directly measured, by Hi-C imaging. Moreover, by defining several distance maps we are able to study  ``averaging'' artifacts that can arise due to various limitations of Hi-C imaging. For example, Hi-C imaging obtains  measurements that are typically averaged across a large heterogeneous distribution of cells that are subjected to nonidentical conditions and exist at nonidentical states in their cell cycles.

\subsection*{Microscope Image Acquisition, Processing and Analysis}
\subsubsection*{Yeast Strains for Experiment}
\label{subsec:strain}
The budding yeast strains used in this study were obtained by transforming the yeast strain EMS219 (Mat alpha, his5 leu2-3,212 ura3-50 CAN1 asp5 gal2 (form I1 rDNA::leu2 URA3+)) with CDC14-GFP:KAN\textsuperscript{R}, to label the nucleolus, and SPC29-RFP:HYG\textsuperscript{R}, to label the spindle pole body, to generate the yeast strain DCY1021.1. DCY1021.1 was transformed to knock out FOB1 and HMO1 to generates DCY1055.1 and DCY1056.1 respectively.

\subsubsection*{Image Acquisition and Baseline Processing}
\label{subsec:image}
Fluorescent image stacks of unbudded yeast cells were acquired using a Eclipse Ti wide-field inverted microscope (Nikon) with a 100$\times$ Apo TIRF 1.49 NA objective (Nikon) and Clara charge-coupled device camera (Andor) using Nikon NIS Elements imaging software (Nikon). Each image stack contained 7 Z-planes with 200 nm step-size.

Image stacks of experimental images were cropped to 7 Z-plane image stacks of single cells using ImageJ and saved as TIFF files. The cropped Z-stacks were read into MATLAB 2018b (MathWorks), converted into maximum intensity projections, and the projections of hmo1$\Delta$ and fob1$\Delta$ were cropped to $55 \times 55$ pixels, to match the dimensions of WT projections, using MATLAB function padarray with \texttt{replicate} option specified to extend outer edge of pixel values to ensure the center of all cropped images was the brightest pixel. The intensity values all projections were normalized by subtracting all intensity values by the minimum value and then dividing the resulting values by the maximum intensity value after subtraction. The normalized intensity values were stored with double point precision, preventing any loss in dynamic range. 

\subsubsection*{Image Analysis}
\label{subsec:image2}
The areas of nucleolar signals were determined by setting all values below threshold, calculated using \texttt{multithresh} function, to NaN and then summing number of values that were not NaNs. That pixel count was converted to $\mu m^2$ by multiplying the sums by $0.0648^2$, the area of each pixel in $\mu m^2$. 

To calculate the standard deviation of the intensities of the nucleolar signal, the non-NaN values remaining after thresholding were re-normalized using the same method described above, and the standard deviation of those values was measured. 

To count clusters within the nucleolar signal, the normalized images were deconvolved with $5\times 5$ Gaussian structural element, using deconvblind function, and underwent two rounds of background subtraction by setting all intensity values below threshold value, calculated using default \texttt{multithresh} function, to NaN and then all NaN values to 0. Clusters were identified using the \texttt{imregionalmax} function and counted using \texttt{bwconncomp} function. 

The simulated images generated from our simulations were analyzed as described above with the additional step of measuring the standard deviation of the each simulated maximum intensity projection. All WT images were analyzed using the script wtExpIm.m. All hmo1$\Delta$ and fob1$\Delta$ images were analyzed using the script cropExpIm.m. All simulated images were analyzed using the script clusterCountLoop.m.

All MATLAB scripts have been made available at \cite{bloom_code}.

\subsection*{Cluster Identification via Network Community Detection}
\label{sec:com}

\subsubsection*{Gene-Interaction Networks from Pairwise-Distance Data}\label{sec:meth_network}

Given a pairwise distance map $X\in\mathbb{R}^{N \times N}$ in which each entry $X_{ij}$ gives the (possibly averaged) Euclidean distance between beads $i$ and $j$ as described in \nameref{sec:meth_distance}, we construct a network model in which there are weighted edges (i.e., interactions) only between beads that are in close proximity to each other and for which each edge weight $A_{ij}\ge 0$  decreases monotonically with distance $X_{ij}$. We propose a model with two parameters, $d^*$ and $s$, which represent a distance threshold and a decay rate, respectively. In particular, we define a network adjacency matrix $A$ having entries
\begin{equation}
A_{ij} =  \left\{ \begin{array}{rl}e^{ -sX_{ij} },& {X_{ij}<d^*}\\ 0,&{X_{ij}\ge d^*}. \end{array}\right.\label{eq:net_map}
\end{equation}
Note that there exists an undirected edge between $i$ and $j$ (i.e., $A_{ij}>0$) only when $X_{ij}<d^*$, and $s$ controls the rate in which the edge weight $A_{ij}$ decreases with increasing distance $X_{ij}$.  \cref{eq:net_map} defines a map between a distance matrix and an affinity matrix that encodes a network.
Note that for any such adjacency matrix $A$, we can equivalently define the network using the graph-theoretic formulation $G(\mathcal{V},\mathcal{E})$. Here $\mathcal{V}=\{1,\dots,N\}$ denotes the set of nodes (i.e., the beads in the chromosome model) and  $\mathcal{E}=\{(i,j,A_{ij}): A_{ij}>0\}$ denotes the set of weighted edges (i.e., a set encoding which beads are interacting as well as their interaction strengths).

In \nameref{sec:meth_distance} we defined several versions of pairwise distance maps---instantaneous, time-averaged, and population-averaged maps---and a network model can be constructed for any of these maps:
\begin{itemize}
\item An \emph{instantaneous interaction network} refers to a network associated with an instantaneous pairwise distance map $X(t)$. 
\item A \emph{time-averaged interaction network} refers to a network associated with a time-averaged pairwise distance map $Y(\tau)$. We point out that due to the nonlinearity of \cref{eq:net_map}, a network associated with a time-averaged distance map can in general differ from a temporal average of instantaneous interaction networks, averaged across the same time interval.
\item A \emph{population-averaged interaction network} refers to a network associated with an population-averaged distance map $\langle Z(t) \rangle_p$. 
\end{itemize}

In addition to the above network models, we are particularly interested in constructing and studying \emph{temporal interaction networks}, which we define as a sequence of time-averaged interaction networks. In particular, given a sequence of timesteps $t\in\{0,1,2,\dots,T\}$, we partition  time into a sequence of  time windows $ \tau(s) = \{(s-1)\Delta + 1,\dots, s\Delta \}$ for $s=1,2,\dots$ of width $\Delta$. We then define a sequence of time-averaged networks $\{G(s)\}$ for $s=1,2,\dots$  associated with these distance maps, which are time-averaged across the nonoverlapping time windows $\{\tau(s)\}$.
In practice, we choose the time window width $\Delta$ to be similar to---but slightly larger than--- $\mu$. Matching the time-scales of  $\mu$ and $\Delta$ allows the temporal interaction networks to efficiently capture the dynamics of interactions. Specifically, if $\Delta$ is too short then the temporal network will be identical across many time steps, which is not an efficient use of computer memory. Moreover, if $\Delta$ is too large, then the temporal network data will be too coarse to identify interaction dynamics occurring at a faster time scale.

Note that this network-construction method involves two parameters: 
the distance cutoff distance $d^*$ and
time-window size $\Delta$.
%and number of window which correspond to the coarsening of time. 
Because the edge weights exponentially decrease with distance, the community detection algorithms we study are insensitive to the choice for $d^*$, provided that $d^*$ is sufficiently large so that the network is connected. Our choice $d^*=325$ in Section: \nameref{sec:stability} ensures there is an edge between all beads in the same cluster (recall Fig.~6 in the main text) and was found to yield qualitatively similar results for other choices of $d^*$.
We also chose $\Delta=10$ to aggregate the time-varying bead-location data (which was saved every 0.1 second) into 1-second intervals. We studied 1000 such time windows to produce \cref{fig:persistence1}.

%In producing the multilayer network, we combined the simulation time steps into "time windows" such that each window produces a single network layer. For our community detection results in this paper, we used 1000 windows, each containing 10 timesteps (1 second of simulation time).

\subsubsection*{Spatiotemporal Gene Clusters Revealed by  Community Detection in Temporal Networks}\label{sec:multilayer_modularity}

We analyze spatiotemporal clustering of chromosomes using community detection methodology for temporal  interaction networks, particularly an approach based on multilayer-modularity optimization. We study the \emph{multilayer modularity measure} \cite{mucha2010community}
\begin{equation}
Q = \frac{1}{2\mu} \sum_{i,j,s,r} \left[ \left(A_{ij}^s - \gamma \frac{k_i^s k_j^s}{2m_s}\right)\delta(s,r) + \omega \delta(i,j) C_{sr} \right] \delta(c_{is},c_{jr}) ,
\end{equation}
where $A_{ij}^s$ denotes the adjacency matrix for network layer $s$ (i.e., that associated with time window $\tau_s$), 
 $\gamma$ is again a tunable ``resolution parameter,'' $k_i^s=\sum_j A_{ij}^s$ is the weighted node degree for  node $i$ in layer $s$, $2m_s = \sum_{ij} A_{ij}^s $ is twice the total number of edges in layer $s$, $\delta({m,n})$ is again a Dirac delta function, $C_{sr}=\delta(s,r-1) + \delta(s,r+1) $ defines the coupling between  consecutive (time) layers and $C_{sr}=1$ if only if $r=s\pm 1$ (otherwise $C_{sr}=0$), and $\{c_{is}\}$ are the integer indices that indicate the community for each node $i$ in each layer $s$. If one wished to analyze a just a single network (e.g., a time-averaged or population-averaged network), then one can simply set $C_{sr}=0$ so that the second term in the square brackets is discarded.

Note that multilayer modularity involves two parameters $\gamma$ and $\omega$, which are ``tuning knobs'' to identify clusters in which their size and temporal coherence are appropriate. In practice, we explore a wide range of
parameter values $\gamma \in[\gamma_{\min} ,\gamma_{\max}]$ and $\omega\in[\omega_{\min},\omega_{\max}]$ to study the multiscale organization of clusters. This approach efficiently explores clustering phenomena at multiple spatial and temporal scales, identifying at which scales clustering is most prevalent and at which scales clustering is nonexistent. 
%In the Supplementary Material: S1 Text we %
To identify appropriate values for $\gamma$ and $\omega$, we used a variety of techniques including the CHAMP algorithm~\cite{weir2017post} (which utilizes fast algorithms that detect convex hulls in the $(\gamma,\omega)$ parameter space) and comparisons to other community-detection algorithms including the study of connected-components.

\begin{comment}
\section*{Additional information}
\subsection*{Funding}
DT was funded by Simons Foundation Award Number 578333. 
MGF acknowledges support from NSF 
DMS-1462992, 
DMS-1517274, 
DMS-1664645 and
DMS-1816630.
BW was supported by NIH 1 T32CA201159-01.
CH acknowledges support from 
NSF DMS-1517274 and NIH U01HL131072.

\subsection*{Author Contributions}
BW and DT developed the community detection algorithms. BW, CH, and DA performed the chromosome model simulations.  MGF, BW, and DT designed the study, analyzed and interpreted the results, and together with CH wrote the manuscript.

\subsection*{Author ORCIDs}
Ben Walker, http://orcid.org/0000-0002-1618-4356 \\
Dane Taylor, http://orcid.org/0000-0003-1851-3309\\
Caitlin Hult, http://orcid.org/0000-0002-2641-2180\\
David Adalsteinsson, http://orcid.org/0000-0002-6767-6713\\
M. Gregory Forest, http://orcid.org/0000-0002-7718-4456\\
\end{comment}

%\section*{Acknowledgements}
%The authors acknowledge motivation for this study as well as valuable discussions with Kerry Bloom, Josh Lawrimore, and Paula Vasquez.

%\section*{Competing Interests}
%The authors declare no competing interests.

%%%%%%%%%%%%%%%%%%%%%%%%%%%%%%%%%%%%%%%%%%%%%%%%
%%%%%%%%%%%%%%%%%%%%%%%%%%%%%%%%%%%%%%%%%%%%%%%%
\clearpage
\section*{Supporting Information}\label{sec:SI}

% \paragraph*{S1 Text.}
% \label{S1_Text.}
% Supporting document that describes the techniques that we use to identify parameters for the multiscale-modularity community-detection algorithm \cite{mucha2010community} that yield clusters having the appropriate spatial and temporal scales.

\paragraph*{S1 Video.}
\label{S2_Video}
One minute of real-time simulation for $\mu=0.09$, demonstrating association of beads into large, stable clusters. Resource available at \url{https://github.com/bwalker1/chromosome-videos/blob/master/Dataset0_allRed_finer_realtime.mp4}.

\paragraph*{S2 Video.}
\label{S3_Video}
One minute of real-time simulation for $\mu=0.19$, showing association of beads into clusters that exhibit frequent changes in membership. Resource available at \url{https://github.com/bwalker1/chromosome-videos/blob/master/Dataset6_allRed_finer_realtime.mp4}. 

\paragraph*{S3 Video.}
\label{S4_Video}
 One minute of real-time simulation for $\mu=1.6$, showing no apparent structure in the bead positions beyond simple pairing. Resource available at \url{https://github.com/bwalker1/chromosome-videos/blob/master/Dataset12_allRed_finer_realtime.mp4}.

\paragraph*{S4 Video.}
\label{S5_Video}
 One minute of real-time simulation for the rigid clustering regime with $\mu=0.09$. We observe that the stable and well-separated clusters have been distinctly labeled by the community detection algorithm. Resource available at \url{https://github.com/bwalker1/chromosome-videos/blob/master/Dataset0_color_finer_realtime_altView.mp4}.

\paragraph*{S5 Video.}
\label{S6_Video}
One minute of real-time simulation for the flexible clustering regime with $\mu=0.19$. Here the clusters are not so clearly separated, but the colored labels still appear consistent with what one would expect. Resource available at \url{https://github.com/bwalker1/chromosome-videos/blob/master/Dataset6_color_finer_realtime_altView.mp4}.

\paragraph*{S6 Video.}
\label{S7_Video}
One minute of real-time simulation for the non-clustering regime with $\mu=1.6$. Here there are no clusters present in the data, but the community detection algorithm still tries to give the same label to nearby beads. Due to the lack of stable community structure, beads change label more frequently than for smaller values of $\mu$. Resource available at \url{https://github.com/bwalker1/chromosome-videos/blob/master/Dataset12_color_finer_realtime_altView.mp4}.

\end{document}